\documentclass[%
apl,twocolumn,showpacs,superscriptaddress,letterpaper]{revtex4-2}

\usepackage{graphicx}
\usepackage{dcolumn}
\usepackage{bm}
\usepackage{hyperref}
\usepackage[utf8]{inputenc}
\usepackage[T1]{fontenc}
\usepackage{mathptmx}
\usepackage{etoolbox}

\makeatother

\usepackage[dvipsnames]{xcolor}
\definecolor{Mycolor2}{HTML}{C2185B}

\hypersetup{
	colorlinks=true,
	linkcolor=blue,
	filecolor=magenta,
	urlcolor=blue,
	citecolor= blue,
}

\usepackage[T1]{fontenc}

\makeatother
\begin{document}
	
\title{Electron-phonon coupling, critical temperatures and gaps in $\rm{NbSe_2}$/$\rm{MoS_2}$ Ising superconductors}

	\date{\today}
	\author{Shubham Patel}\email{spatelphy@iitkgp.ac.in}
	\affiliation{Department of Physics, Indian Institute of Technology, Kharagpur-721302, India}
	\author{Soumyasree Jena}
	\affiliation{Department of Chemistry, Indian Institute of Technology, Madras-600036, India}
	\author{A Taraphder$^1$}\email{arghya@phy.iitkgp.ac.in}
	
	\begin{abstract}
		Utilizing Migdal-Eliashberg theory of superconductivity within the first-principles calculations, we work out the role of electron-phonon coupling (EPC) and anisotropic superconducting properties of a recently discovered [PRB 104, 174510 (2021)] 2D van der Waals heterostructure comprising a single layer of MoS$_2$ and few layers of NbSe$_2$. We find strong EPC and a softening of phonon modes in the lowest acoustic branch. While the single MoS$_2$ layer does not actively contribute to the EPC, it significantly elevates the superconducting critical temperature ($T_c$) compared to monolayer NbSe$_2$. This is attributed to the degradation of the charge-density wave (CDW) by the MoS$_2$ layer. Notably, we observe a two-gap superconductivity in $\rm{NbSe_2}$/$\rm{MoS_2}$ and extend our study to three layers of NbSe$_2$. A reduction in $T_c$ with increasing thickness of NbSe$_2$ is observed. We confirm that this trend is consistent with recent experiments, if one goes beyond three layers of NbSe$_2$. Incorporation of spin-orbit coupling (SOC) suggests a possible mechanism for Ising superconductivity. We find that SOC reduces EPC while $T_c$ is suppressed concomitantly by about 5K, leading to a closer estimate of the experimental $T_c$.
	\end{abstract}
	
	\maketitle
	
	Charge density wave (CDW) and superconductivity (SC) are competing instabilities in a vast variety of materials such as TMDs \cite{ugeda2016characterization, koley2020charge}, intercalated materials, and recently investigated Kagome metals \cite{yu2021unusual}. TMDs have drawn great interest lately due to their device application potential. Apart from their semiconducting, spintronic and valleytronic attributes, TMDs are studied for the competition between CDW and SC \cite{gruner1988dynamics,borisenko2009two,thomson1994scanning,weber2011extended,dordevic2003optical,taraphder2011preformed,koley2014preformed}. Other than disorder, dimensionality also plays a significant role in manipulating the behaviour of these systems. \cite{calandra2009effect,lian2023interplay}. 
	
	Van der Waals heterostructures have captured considerable attention over the past two decades since the discovery of graphene due to the display of unique properties that hold immense promise for electronic and optoelectronic applications. The prospect of crafting a van der Waals heterostructure with intriguing emergent properties appears to be a novel route towards advanced tunable devices. In particular, the transition metal dichalcogenides (TMDs), which are categorized as two-dimensional semiconductors, emerged as noteworthy candidates. Among these, $\rm{MoS_2}$ has garnered substantial interest as a potential semiconductor, showcasing Ising superconductivity (ISC) in few-layer samples when subjected to electrostatic gating \cite{lu2015evidence}. Likewise, $\rm{NbSe_2}$ \cite{xi2016ising} represents another potential Ising superconductor. $\rm{NbSe_2}$ and $\rm{TaS_2/Se_2}$ had a long-standing appeal to experimental and theoretical investigators owing to the possible co-existence of CDW and SC \cite{freitas2016strong,koley2020charge}. Along with CDW and SC, ferromagnetism is also reported in this material \cite{zhu2016signature}, which makes it even more intriguing.
	
	$\rm{2H-NbSe_2}$ is a well-studied TMD material in which CDW and SC coexist at lower temperatures (CDW at 33K with coexisting SC below 7K) \cite{xi2015strongly}. Electron-phonon coupling (EPC), along with electronic correlation \cite{dordevic2003optical}, plays a crucial role in determining its instabilities rather than Fermi surface (FS) nesting \cite{calandra2009effect}. The interpocket and intrapocket scatterings across the FS are proposed mechanisms for CDW and SC, respectively, in the monolayer system \cite{zheng2019electron}. $\rm{2H-MoS_2}$ is a very well-known member in the TMD-family that has been studied quite extensively due to its large spin-orbit coupling \cite{zhu2011giant} and several other topological, electronic, optical, and catalytic properties. 
	
	It is found experimentally that 2D samples of $\rm{NbSe_2}$ are unstable at ambient conditions. Theoretically also, the high symmetry phase of $\rm{NbSe_2}$ is predicted to be unstable. There is a transition to a distorted CDW $3\times3$ superstructure \cite{calandra2009effect} as the temperature is reduced. In a recent study, ISC is claimed in the heterostacking of a few layers of $\rm{NbSe_2}$ with a single layer of $\rm{MoS_2}$, where the authors argue in favour of the stability of a 2D SC \cite{baidya2021transition} state. Though the competition between CDW and SC was not addressed, it seems $\rm{MoS_2}$ plays a very crucial role in mitigating the effects of CDW, stabilizing the SC state by preventing a clustering of $\rm{NbSe_2}$ into a CDW superstructure, reported in an earlier study on intercalated $\rm{NbSe_2}$ bilayers \cite{yin2023quenched}.
	
	ISC has gained prominence as a burgeoning field in the search for unconventional SC lately \cite{lu2015evidence,xi2016ising,de2018tuning,zhou2016ising}. ISC primarily originates in systems lacking inversion symmetry that leads to an intrinsic SOC. TMDs are the main candidates in which ISC is likely to occur. In these systems, spins align themselves in the out-of-plane direction, and it takes a high in-plane upper critical field to destroy the superconductivity. For these superconductors, the in-plane magnetic field crosses the Chandrasekhar-Clogston-Pauli limit, $B_p = 1.86\,T_c$. Therefore, a considerable influence of SOC on the superconducting properties of Ising superconductors is anticipated. But SOC is not the only factor which affects ISC, spin-orbit scattering \cite{coleman1983dimensional}, singlet-triplet mixing \cite{wickramaratne2020ising}, spin-fluctuations \cite{das2023electron}, disorder \cite{mockli2020ising} and intervalley scattering \cite{ilic2017enhancement} could affect ISC concurrently. In this work, we will also discuss the effect of SOC on electronic and superconducting properties.

\begin{figure}[!htb]
		\centering
		\begin{tabular}{l}
			\includegraphics[scale=0.15]{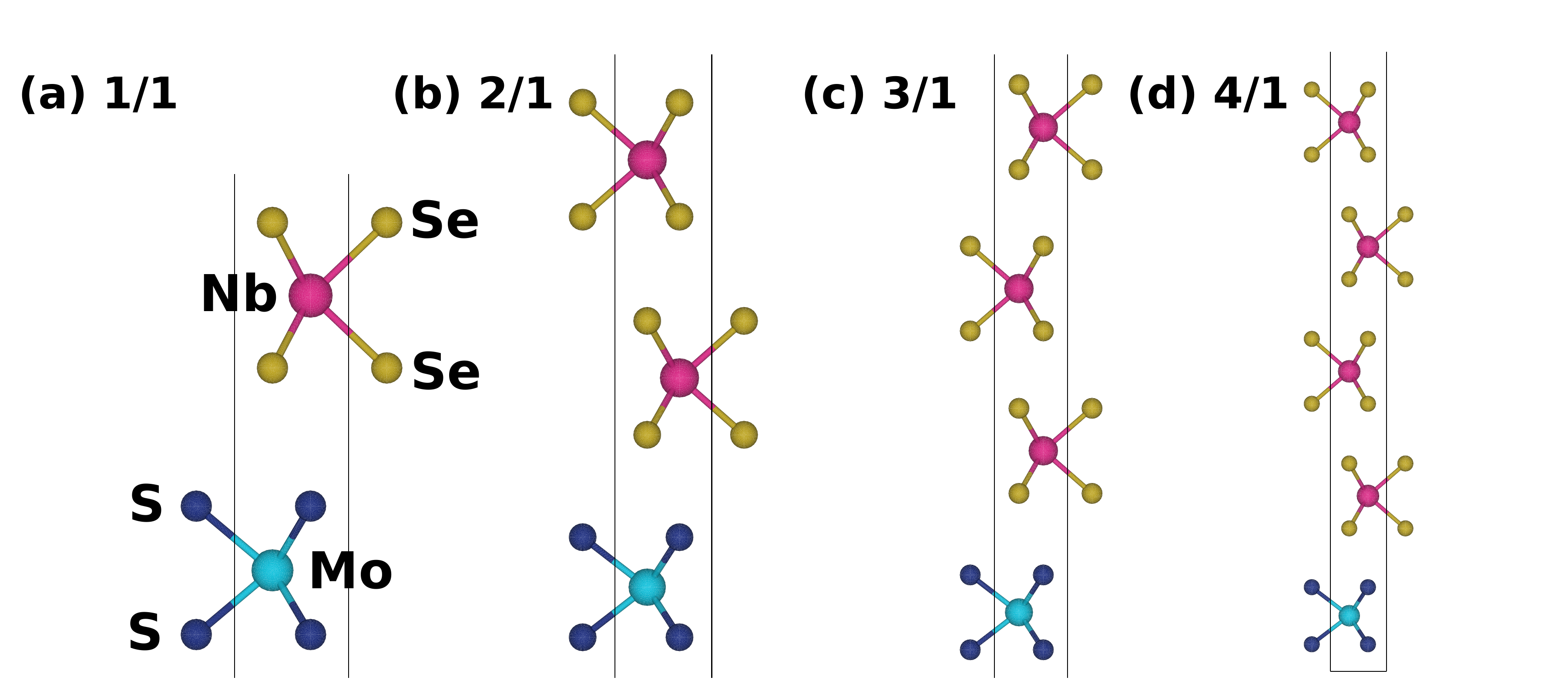}
		\end{tabular}
		\caption{The crystal structures of $\rm{NbSe_2}$/$\rm{MoS_2}$ (a) 1/1, (b) 2/1, (c) 3/1 and (d) 4/1 heterostructures. In all four cases, a single layer of $\rm{MoS_2}$ and various layers of $\rm{NbSe_2}$ are used. Magenta, yellow, cyan and blue spheres, in the left figure, represent Nb, Se, Mo and S atoms, respectively. A vacuum of 20 \AA\, is introduced in all the geometries along the c-axis.}
		\label{fig:struct}
	\end{figure}

	2H polymorphs of $\rm{MoS_2}$ and $\rm{NbSe_2}$ are members of transition metal dichalcogenides (TMDs) family with $D_{6h}$ point group symmetry in which transition metals are sandwiched between the chalcogen layers. We model 2D heterostructures with one layer of $\rm{MoS_2}$ and $n$-layers of $\rm{NbSe_2}$ ($n/1$ configuration). $\rm{MoS_2}$ with lattice constant ($a$) of $3.19$ \AA\, works as a substrate which produces $6$\% (for 1/1) of tensile strain while stacked with $\rm{NbSe_2}$ ($a = 3.39$ \AA). The in-plane lattice constants for different layers of $\rm{(NbSe_2)_n}/\rm{MoS_2}$ are provided in Table T1 of supplementary material (SM) \cite{patel2024electron}. The negative binding energy [$E_b(eV) = E_{\rm{NbSe_2}/\rm{MoS_2}} - E_{\rm{NbSe_2}} - E_{\rm{MoS_2}}$] implies a strong possibility of the formation of $\rm{NbSe_2}$/$\rm{MoS_2}$ heterostructures as shown in Table T1 of the SM \cite{patel2024electron}. In the present work, we go up to four layers of $\rm{NbSe_2}$ that are stacked with $\rm{MoS_2}$ in a particular stacking with respect to each other as shown in Fig. \ref{fig:struct}. It is important to mention that the AB-stacking is used to form the crystal structures, which is found to be the most stable stacking for these hexagonal heterostructures \cite{patel2022electric}. In AB-stacking, the chalcogen atom and the transition metal atom from different layers are on top of each other vertically along $c-$axis. The monolayers of $\rm{MoS_2}$ and $\rm{NbSe_2}$ show semiconducting and metallic properties individually. $\rm{NbSe_2}$ undergoes a structural transition to CDW phase at finite temperature (33K) and also shows superconducting properties at even lower temperature around 7K \cite{zheng2019electron}.
	
	Electronic structure calculations show (See Fig. S1 in the SM \cite{patel2024electron}) that $d$ orbitals of $\rm{Nb}$ atoms mainly contribute at the Fermi level (FL). The topology of the $\rm{Nb}$ bands is nearly intact close to the FL and remains the same as for the $\rm{NbSe_2}$ monolayer. Also, the overall band structure from $\rm{MoS_2}$ layer is similar to the pristine monolayer, except the conduction bands are shifted downward and interact with the $\rm{Nb}$ bands, which clearly indicates interfacial interactions between the two layers at the interface. In the valence band region, one may notice a hybridization between $\rm{Mo}$ $d$ orbitals and $\rm{Se}$  $p$ orbitals. In the FS plot there are two types of hole pockets, one at the $\Gamma$ and the other at the $K$ point. The emergence of the hole pocket at the $\Gamma$ point is attributed to Nb $d_{x^2-y^2}$ and $d_{xy}$ orbitals, whereas the hole pocket at the $K$ point arises from the Nb $d_{z^2}$ orbital. Increasing the number of layers of $\rm{NbSe_2}$ has no significant effect on the $\rm{MoS_2}$ bands except shifting the MoS$_2$-derived bands downward as they cross the FL eventually for $n > 4$. The reason for this downward shift is the tensile strain induced in the heterostructures with increasing layers \cite{patel2022electric}. On the other hand, it is obvious that on increasing $n$, more bands populate the FL. The lowest band, which is closer to FL and belongs to the lowest NbSe$_2$ layer, comes closer to FL and becomes flatter, increasing the size of the hole pockets at the $\Gamma$ point.

	We also perform calculations including spin-orbit coupling (SOC). The SOC bands are shown in Fig. S1(c), where one may notice that SOC has a significant effect on $\rm{NbSe_2}$ bands in all the cases. Despite the significant spin-splitting observed in the bands of $\rm{MoS_2}$, its considerable distance from the FL implies the splitting is not going to alter the physical picture substantially. At the $K$ point, the $\rm{NbSe_2}$ bands at FL have a noticeable Zeeman type spin-splitting. A negligible Rashba spin-splitting can also be seen at the valence band maxima at $\Gamma$ point in all the cases. This is similar to the case of monolayer $\rm{MoS_2}$. This Rashba type spin-splitting in the bands near the interface is the signature of 2D electron-gas at the interface of $\rm{NbSe_2}$ and $\rm{MoS_2}$. The presence of SOC in van der Waals materials is pivotal in influencing electron-phonon interactions, which will be discussed later.

	\begin{figure}[!htb]
		\centering
		\begin{tabular}{l}
			\includegraphics[scale=0.52]{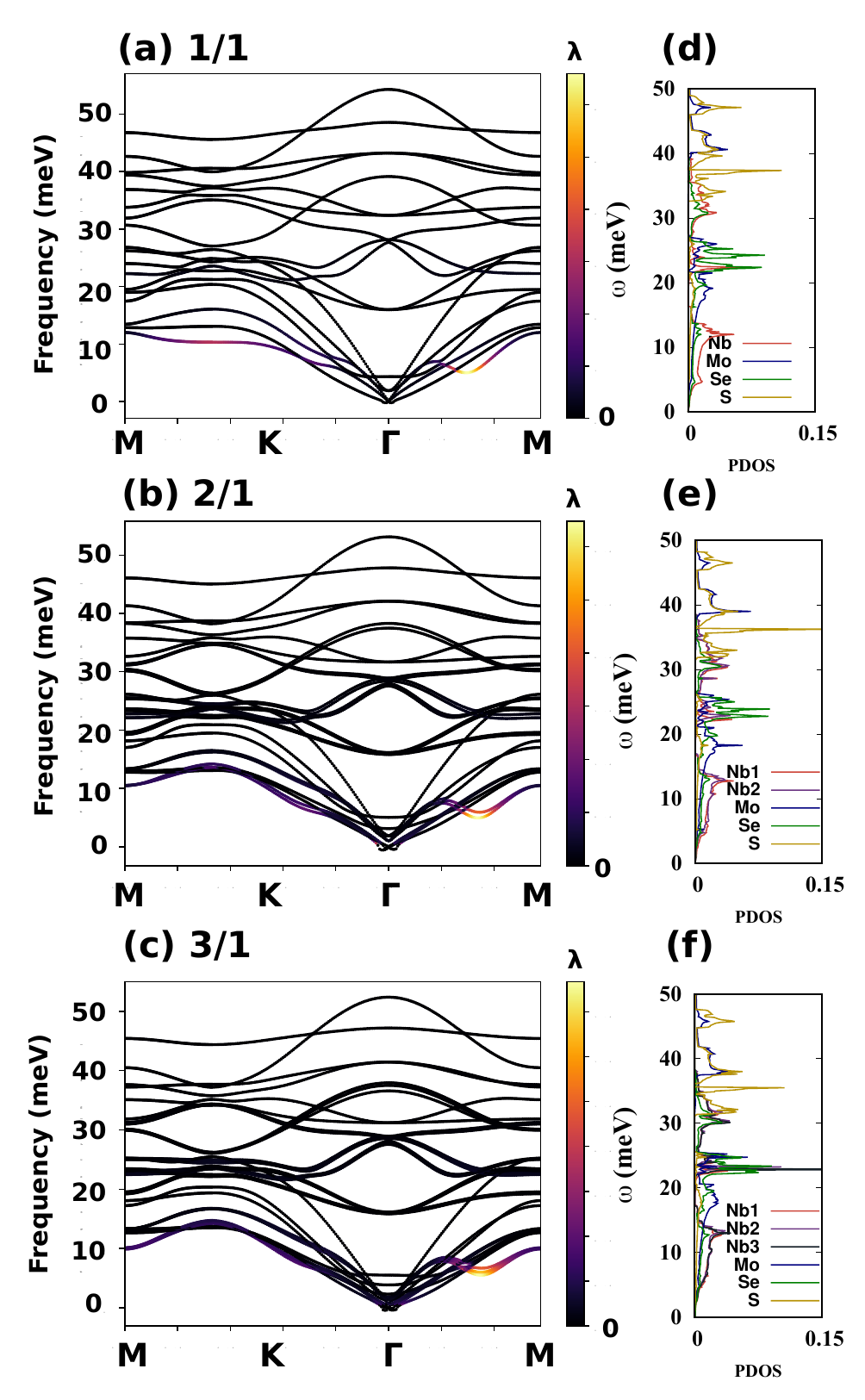}
		\end{tabular}
		\caption{(a-c) The phonon spectra and mode-resolved EPC for all three configurations. Absence of imaginary frequency in the phonon spectra implies stability of the heterostructures. The negligible softening around the $\Gamma$ point is due to numerical artefact (see text). The softening along the $\Gamma-M$ indicates a suppressed CDW. The color scale indicates the strength of EPC. (d-f) The partial DOS in the right panels.}
		\label{fig:phonon}
	\end{figure}

\begin{figure*}[!htb]
		\centering
		\includegraphics[scale=0.75]{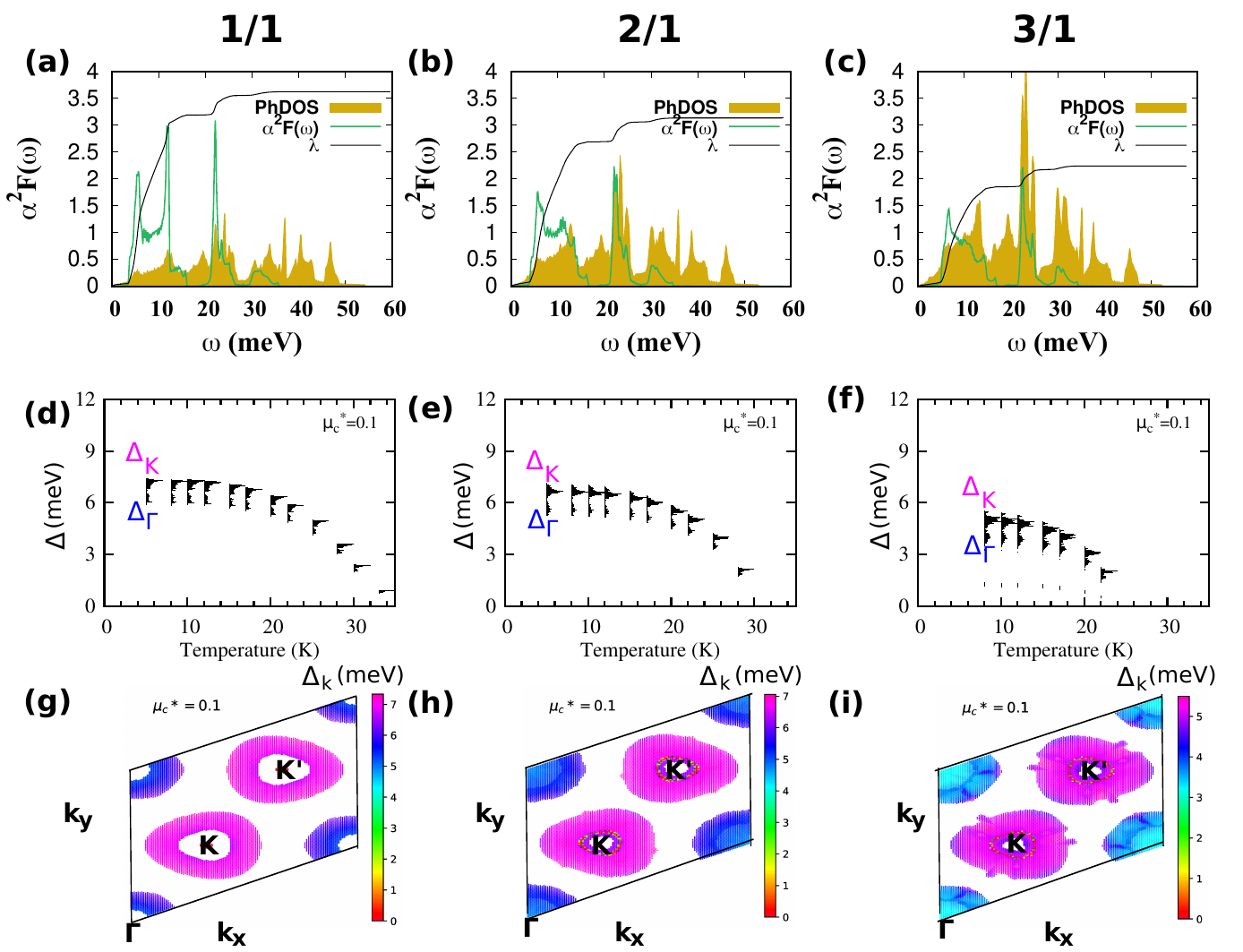}
		\caption{The Eliashberg spectral function ($\alpha^2F(\omega)$, green) along with the integrated EPC strength ($\lambda$, black) for (a) 1/1, (b) 2/1 and (c) 3/1  structures. The phonon DOS is in yellow. (d-f) The energy distribution of the gap ($\Delta$) as a function of temperature at $\mu_c^* = 0.1$. $\Delta_{K}$ and the $\Delta_{\Gamma}$ are the gap distributions around the K and $\Gamma$ points, respectively. The two-gap nature is clear from the gap-resolved FS (g-i) plotted at $T = 8$K; the nature remains the same for other temperatures.} 
		\label{fig:a2f}
	\end{figure*}

	To calculate the phonon-mediated superconducting properties of the layered $\rm{NbSe_2}$/$\rm{MoS_2}$ heterostructure, we first calculate the dynamical matrices using density functional perturbation theory as implemented in QE.  There is no visible gap in the spectrum over the whole frequency range. However, from the partial density of states it is clear that the lowest few phonon modes (acoustic) are contributed by Nb (the red curve in the phonon DOS for 1/1 of Fig. \ref{fig:phonon}(d)), followed by a gap in the range between $15$ to $20$ meV for the Nb modes. All the modes coming from Nb in systems with $n > 1$ are equally populated in the acoustic phonon region in the range from $0$ to $15$ meV, as shown in Figs. \ref{fig:phonon}(a-c). There is a finite but very small contribution of Mo atom to the acoustic modes in comparison to contributions from NbSe$_2$. The optical branches are coming mainly from S, and the Mo atom has a substantial contribution in the higher frequency regime. There is a small hybridization of Se with Nb in the acoustic modes, which suggests that the NbSe$_2$ layers play a substantial role in EPC. One can also note that beyond $35$ meV, neither Nb nor Se contributes to the optical branches, and it will be more explicit in the spectral function plot shown in Fig. \ref{fig:a2f}, that, beyond $35$ meV, $\alpha^2F(\omega)$ vanishes. This implies NbSe$_2$ layers only are responsible for EPC and electron-phonon induced SC. The plots of mode-resolved $\alpha^2F(\omega)$ in the SM vindicates this clearly. These plots and details thereof are provided in Fig. S1 of SM \cite{patel2024electron}. To be sure of the source of EPC, mode-resolved electron-phonon linewidth plots are also shown in Fig. \ref{fig:phonon}, using the relation \cite{margine2013anisotropic,lee2023electron},
	$$	\lambda_{qv} = \frac{1}{N(\epsilon_F)\omega_{qv}}\sum_{nm}^{}\int_{BZ}^{} \frac{dk}{\Omega_{BZ}}|g_{mn,v}(k,q)|^2 $$
	\begin{equation}
		\times \delta(\epsilon_{nk}-\epsilon_F)\delta(\epsilon_{mk+q}-\epsilon_F).
	\end{equation}
	The parameters are defined in the computational details section of the SM \cite{patel2024electron}. One can clearly observe a phonon-softening in between $\Gamma$ and $M$ high-symmetry points in the LA phonon branch, which is known to be $E'$ vibrational mode of the acoustic branch. This mode belongs to the NbSe$_2$ layer and indicates the movement of Nb and two Se atoms in the same in-plane direction. It is interesting to point out that in monolayer NbSe$_2$, this phonon-softening occurs at a \textbf{q} point where, \textbf{q} $ = \frac{2}{3}\Gamma M$ \cite{zheng2019electron,lian2022intrinsic}, implying a $3\times3$ CDW instability. In the case of 1/1-NbSe$_2$/MoS$_2$, we observe it at \textbf{q} $ = \frac{1}{2}\Gamma M$ (Fig. \ref{fig:phonon}(a)), suggesting a commensurate $2\times2$ CDW ordering. As the number of NbSe$_2$ layers increases, the softening approaches \textbf{q} $ = \frac{2}{3}\Gamma M$ gradually, possibly through a series of incommensurate CDWs, at least for 2/1 and 3/1 configurations. The color scale suggests that almost all the EPC is concentrated in this phonon-softening valley of the acoustic branch of NbSe$_2$, and the other modes do not contribute to EPC. This is true for all the three cases \ref{fig:phonon}(a-c). A similar softening is also found in bare NbSe$_2$ monolayer \cite{zheng2019electron}; the only difference is that the softening in our case is between $\Gamma$ and $M$ points while it is close to the $M$ point in the earlier study, indicating that a single layer of MoS$_2$ affects the EPC strength in NbSe$_2$/MoS$_2$ heterostructures.

	Next we calculate EPC strength, $\lambda$, as a function of frequency and show in Fig \ref{fig:a2f}(a) along with Eliashberg spectral function $\alpha^2F(\omega)$. $\alpha^2F(\omega)$ vanishes beyond $35$ meV. Comparison of $\alpha^2F(\omega)$ with partial DOS (in Fig. \ref{fig:phonon}(d-f)) shows that the contribution from NbSe$_2$ layers also vanishes beyond this range; that means NbSe$_2$ layers are sole contributors to EPC. The $\alpha^2F(\omega)$ also behaves in the same manner as the partial DOS in the range $0-35$ meV. The gaps in $\alpha^2F(\omega)$ in the range $15-20$ meV and $25-30$ meV in Fig. \ref{fig:a2f}(a-c) are apparent from the partial DOS of Nb and Se in Fig. \ref{fig:phonon}(d-f). The cumulative EPC, indicated by $\lambda$ in Figs. \ref{fig:a2f}(a-c), is calculated using Eqn. (2), described in the SM \cite{patel2024electron}. Clearly, $\lambda$ reduces from $3.52$ to $2.4$ as the NbSe$_2$ layer numbers increase from 1 to 3 (i.e., 1/1 to 3/1 heterostructure) respectively, due to the reduction of $\alpha^2F(\omega)$. This reduction in total EPC will also result in the reduction of $T_c$. A sudden jump in $\lambda$ at a frequency, $\omega = 23$ meV is noticed in all three cases, due to a gap opening in the spectral function between acoustic and optical phonon modes of Nb.

	To evaluate the superconducting $T_c$, fully anisotropic Migdal-Eliashberg theory \cite{ponce2016epw,margine2013anisotropic} is  employed. Fig. \ref{fig:a2f}(d-f) show the energy distribution of the superconducting gap as a function of temperature at an effective Coulomb potential $\mu^* = 0.1$. The leading edge of the gap function ($\Delta_0$) at $T = 0$K is at 6.7 meV. The ordinary Allen-Dynes modified McMillan equation \cite{allen1975transition} gives a superconducting $T_c = 20$K, while anisotropic Migdal-Eliashberg theory gives $T_c = 35$K. This large difference in the two superconducting $T_c$ is a signature of anisotropy in the superconducting gap distribution on the FS, which is a consequence of the multi-sheet nature of FS in NbSe$_2$/MoS$_2$. This was observed \cite{margine2013anisotropic} in MgB$_2$ ($T_c = 39$K) too, which is a two-gap superconductor. In NbSe$_2$/MoS$_2$, there are hole-pockets around $\Gamma$, $K$ and $K'$ valleys in the BZ. The larger gap is associated with the out-of-plane $d_{z^2}$ Fermi-sheets at the $K(K')$ point, while the smaller gap involves in-plane Nb $d_{x^2-y^2}$ and $d_{xy}$ Fermi-sheets located around the $\Gamma$ point, shown in the gap-resolved FS in Figs. \ref{fig:a2f}(g-i).It is interesting to note that the two-gap feature is robust with increasing number of NbSe$_2$ layers.

	The calculated $T_c$ of NbSe$_2$/MoS$_2$ is large as compared to a single layer of NbSe$_2$. The experimental value is 3.8K for the single layer system \cite{frindt1972superconductivity}, while it is 7K for bulk NbSe$_2$ \cite{luo2017s}. An EPW calculation on a 3x3 CDW supercell reported $T_c$ = 4.4K for a single layer \cite{zheng2019electron}. Thus, our study suggests that a single layer of MoS$_2$ has a substantial impact on SC of 2H-NbSe$_2$. Increasing the number of NbSe$_2$ layers up to three shows a constant reduction in $T_c$ in Figs. \ref{fig:a2f}(d-f) and in Table \ref{table:muc}. $T_c$ is sensitive to the choice of Coulomb potential ($\mu_c^*$). An increase in $\mu_c^*$ results in a decrease in superconducting $\Delta_0$ and $T_c$ \cite{margine2013anisotropic}. Since there are no available estimates of $\mu_c^*$ to compare with, we have calculated $T_c$ for different values of $\mu_c^*$ (Table. \ref{table:muc}). 

	ISC with $T_c \simeq 6.5$K has been suggested in a single layer of MoS$_2$ stacked with few layers of NbSe$_2$ ($\sim$15 nm, to be exact) \cite{baidya2021transition,baidya2022correlated} recently. A similar $T_c$ is expected with increasing NbSe$_2$ layers. The variation of $T_c$ with the thickness of NbSe$_2$ suggests an increase in NbSe$_2$ layers will result in a suppression of $T_c$ as shown in scanning tunneling microscopy measurements \cite{khestanova2018unusual}. Migdal-Eliashberg theory, incorporating spin-fluctuations, finds this trend for 1H-TaS$_2$ \cite{lian2022intrinsic} and more recently in 2H-NbSe$_2$ \cite{das2023electron}. In 1H-TaS$_2$, it is attributed to an inevitable suppression of Cooper pair density at the superconductor-vacuum interface. Our computational resources make it difficult to go beyond three layers of NbSe$_2$, however the trend is clear.	

\begin{center}
	\begin{table}[!htb]
		\centering
		\setlength{\tabcolsep}{16pt} 
		\renewcommand{\arraystretch}{1.1}
		\caption{\label{table:muc} Variation of $T_c$ with lowest, moderate and highest possible values of Coulomb potential, $\mu_c^*$. $T_c$ reduces with increasing $\mu_c^*$; $\lambda$ is EPC strength.}
		\begin{tabular}{ c c c c}
			\hline
			\hline
			$\mu_c^*$& & $T_c$ (K) &\\
			\hline
			&	1/1 &	2/1 & 3/1 \\
			&	$\lambda=3.52$ &	$\lambda=3.13$ & $\lambda=2.24$ \\
			\hline
			0.05 	& 37.4 & 34.1 & 28.7\\
			0.10 	& 35.3 & 31.0 & 24.0\\
			0.16    & 30.1 & 27.9 & 21.3\\
			\hline
			\hline
		\end{tabular}
	\end{table}
\end{center}
	
	Furthermore, it is important to emphasize that the Migdal-Eliashberg approach results in a fair agreement for a variety of materials, though it overestimates the superconducting gap and $T_c$ \cite{leroux2015strong,wickramaratne2020ising,das2023electron}. The claim is that in NbSe$_2$ and similar materials that show ISC, the overestimation of phonon-induced SC can be mitigated by spin-fluctuations \cite{wickramaratne2020ising,das2023electron}. For a comparison, we perform the EPW calculations for the NbSe$_2$ monolayer and find an overestimated $T_c = 20$K as reported earlier \cite{das2023electron}. The same could be possible for NbSe$_2$/MoS$_2$ heterostructures.

	Finally, speaking of ISC, a mechanism which is controlled by SOC in non-centrosymmetric materials, we infer that the SOC, which has been ignored in electron-phonon calculations so far, can have a huge impact on the EPC strength. Inclusion of SOC weakens the EPC strength almost by a factor of two as reported earlier in the case of CaBi$_2$ \cite{golkab2019electron} and TaS$_2$ \cite{lian2022intrinsic}, close to the experimentally reported values. $\rm{(NbSe_2)_n}/\rm{MoS_2}$ system exhibits a notable SOC as illustrated in Fig. S1(c), which is evident from a Zeeman type of spin-splitting at the $K$ point in the BZ. Considering this, incorporating SOC into calculations may yield a more accurate estimate of $T_c$. Including SOC, the superconducting $T_c$ is indeed reduced by about 5K for 1/1  ($T_c =30$K, $\lambda = 3.32$) and 2/1-NbSe$_2$/MoS$_2$ ($T_c =26.6$K, $\lambda = 2.57$) as shown in Fig. S3 of SM \cite{patel2024electron}. More experimental investigations should provide a better estimate of $T_c$ in $\rm{(NbSe_2)_n}/\rm{MoS_2}$ ($<15nm$) and validate our results.

	To summarize, we investigated electronic and superconducting properties of AB-stacked NbSe$_2$/MoS$_2$ van der Waals heterostructures and studied the behavior of critical temperature $T_c$ with the number of layers of NbSe$_2$, using Migdal-Eliashberg theory. While NbSe$_2$ layers affect the FL considerably, a single layer of MoS$_2$ does not significantly alter the FS topology, at least up to 4/1 configuration. Moreover, the heterostructure shows a robust electron-phonon coupling (EPC) strength. A pronounced phonon softening in the acoustic modes of the NbSe$_2$ layers is observed along the $\Gamma - M$ direction, attributed to movements of Nb and Se atoms within the $ab$-plane. MoS$_2$ layers do not exhibit any EPC, though their inclusion dramatically enhances the $T_c$ when stacked with NbSe$_2$. This enhancement owes its origin to the stabilization of the crystal structure by MoS$_2$, mitigating the CDW instability of NbSe$_2$. Our findings suggest a substantial increase in $T_c$ when a semiconducting 2D material like MoS$_2$ is combined with a superconducting counterpart like NbSe$_2$. We also suggest that increasing the thickness of NbSe$_2$, stacked with a single layer of MoS$_2$, results in an unusual reduction in $T_c$. We anticipate that the experimentally reported $T_c$ could be attained by adding more (beyond 4) layers of NbSe$_2$. Additionally, we observe that the inclusion of spin-orbit coupling (SOC) leads to a marked decrease in EPC, significantly degrading $T_c$. This also points to the possibility of an Ising superconductivity in this system.

	We acknowledge National Supercomputing Mission (NSM) for providing computing resources of 'PARAM Shakti' at IIT Kharagpur, which is implemented by C-DAC and supported by the Ministry of Electronics and Information Technology (MeitY) and Department of Science and Technology (DST), Government of India.\\

	\bibliographystyle{unsrt}
	\bibliography{refs}

\end{document}


\begin{center}
		\textbf{{\Large Supplementary Material}}\\
		\vspace{0.5cm}
		\textbf{{\Large Electron-phonon coupling, critical temperatures and gaps in $\rm{NbSe_2}$/$\rm{MoS_2}$ Ising Superconductors}}\\
		\vspace{0.5cm}
		\textbf{Shubham Patel$^1$, Soumyasree Jena$^2$, A Taraphder$^1$}\\
		
		$^1$\textit{Department of Physics, Indian Institute of Technology, Kharagpur-721302, India}
		
		$^2$\textit{Department of Chemistry, Indian Institute of Technology, Madras-600036, India}
		
	\end{center}
	
	\section{\label{sec:compsec}Computational details and methodology}
	
	The electronic structure calculations are performed with  Perdew-Burke-Ernzerhof exchange-correlation functional  
	\cite{perdew1996generalized} in the framework of generalized gradient approximation and projector augmented wave (PAW) method \cite{blochl1994projector,kresse1999ultrasoft} as implemented in Vienna \textit{ab initio} simulation package {\small (VASP)}. All the lattice geometries are fully relaxed by the conjugate gradient algorithm until the force on each atom is less than 0.01 eV/\AA\, with energy cutoff 500 eV and an energy tolerance criterion of $10^{-6} $eV. A vacuum of 20\AA\, is introduced to forbid the interactions between periodic layers of the heterostructure. The van der Waals (vdW) interactions are employed with a DFT-D2 correction scheme developed by Grimme \cite{grimme2006semiempirical}. The lattice dynamics and electron-phonon coupling (EPC) are calculated within the norm-conserving pseudopotentials for exchange-correlation functional \cite{troullier1991efficient}, as implemented in {\small QUANTUM ESPRESSO} (QE) package \cite{giannozzi2009quantum,giannozzi2017advanced,giannozzi2020quantum}, with a plane-wave energy cutoff of 100 Ry and methfessel-paxton smearing of 0.02 Ry. The phonon dispersion is obtained by Fourier interpolation of the dynamical matrices computed using a $24\times24\times1$ \textbf{k}-point mesh and a $6\times6\times1$ \textbf{q}-point mesh. 
	
	The anisotropic superconducting properties are calculated using fine \textbf{k} and \textbf{q} grids of $120\times120\times1$ and $60\times60\times1$, respectively,  using Migdal-Eliashberg (ME) formalism as implemented in EPW code \cite{giustino2007electron,ponce2016epw,margine2013anisotropic}. For the Wannier interpolation in EPW, we used maximally localized Wannier functions ($p$-orbitals of Se and S, Nb $d_{xy/x^2-y^2}$, $d_{z^2}$ and Mo $d_{xy/x^2-y^2}$, $d_{z^2}$) to describe the electronic structure near the Fermi level for the case of $\rm{NbSe_2}$/$\rm{MoS_2}$. The Matsubara frequency cutoff is set to $0.4\rm{eV}$, which is 10 times larger than the upper limit of the phonon frequency in el-ph calculations. The mathematical and technical details of Migdal-Eliashberg calculations are described extensively by Allen \cite{allen1983theory}, Margine \cite{margine2013anisotropic} and Ponc{\'e} \cite{ponce2016epw} previously. Here, we concentrate on electronic, vibrational and superconducting properties. The Eliashberg electron-phonon spectral function $\alpha^2F(\omega)$ and the cumulative frequency dependence of EPC, $\lambda(\omega)$ can be calculate by 
	$$	\gamma_{qv} = 2\pi\omega_{qv}\sum_{nm}^{}\int_{BZ}^{} \frac{dk}{\Omega_{BZ}}|g_{mn,v}(k,q)|^2 $$
	\begin{equation}
		\label{eq:gamma}
		\times \delta(\epsilon_{nk}-\epsilon_F)\delta(\epsilon_{mk+q}-\epsilon_F).
	\end{equation}
	\begin{center}
		\label{eq:alpha2F}
		\begin{equation}
			\alpha^2F(\omega) = \frac{1}{2\pi N(E_F)}\sum_{qv}^{} \frac{\gamma_{qv}}{\omega_{qv}} \delta(\omega-\omega_{qv})		
		\end{equation}
	\end{center}
	and
	\begin{center}
		\label{eq:lambda}
		\begin{equation}
			\lambda(\omega) = 2\int_{0}^{\omega}\frac{\alpha^2F(\omega)}{\omega}d\omega,	
		\end{equation}
	\end{center}
	respectively, where $\gamma_{qv}$ is the phonon linewidth associated with momentum $q$ and branch index $v$, inverse of which represents the phonon lifetime and that signifies the EPC strength. $\omega_{qv}$ is the phonon frequency and $N(E_F)$ is the electron density of states (DOS) at the Fermi level. The temperature dependent superconducting gap and DOS are calculated using anisotropic Migdal-Eliashberg theory \cite{margine2013anisotropic}. The Allen-Dynes modified McMillan equation \cite{allen1975transition,mcmillan1968transition} is also employed for the calculation of $T_c$, where an effective Coulomb potential $\mu_{c}^*$ is varied. The structures are visualized in {\small VESTA} software \cite{momma2011vesta}.

	\begin{center}
		\begin{table}[!htb]
			
			\setlength{\tabcolsep}{12pt} 
			\renewcommand{\arraystretch}{1.0}
			\textbf{TABLE T1} The in-plane lattice constants ($a/b$) and binding energies ($E_b$) are tabulated.
			\centering
			\begin{tabular}{ c c c c c}
				\hline
				\hline
				&	1/1  &	2/1   & 3/1    & 4/1\\
				\hline
				$a$ (\AA) 	    & 3.316  & 3.361  & 3.381  & 3.397\\
				$E_b(eV)$ 	& -0.205 & -0.210 & -0.207 & -0.212\\
				\hline
				\hline
			\end{tabular}
		\end{table}
	\end{center}
	
	\section{\label{sec:elec}Electronic structures}
	The band structures, without (Fig. S1(a)) and with (Fig. S1(c)) SOC are calculated for $(\rm{NbSe_2})_n$/$\rm{MoS_2}$ van der Waals heterostructure where, $n = 1-4$. 
	\begin{figure*}[!htb]
		
		\begin{tabular}{l}
			\includegraphics[scale=0.85]{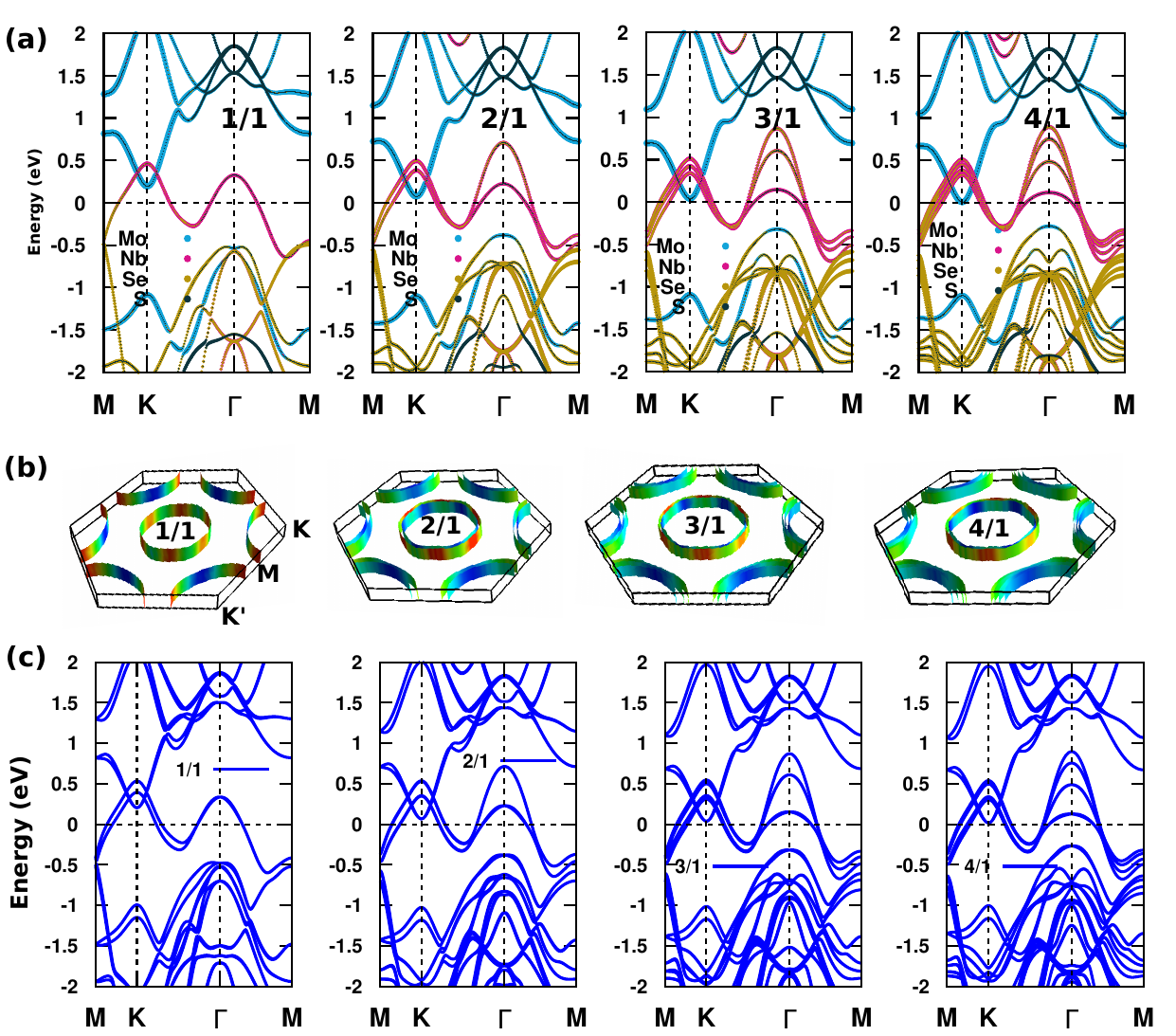}
		\end{tabular}
		\textbf{Fig. S1} Atom-projected band structures for the four heterostructures where the number of $\rm{NbSe_2}$ layers increases from one to four from left to right (upper panel). Fermi surface sheets are shown in the middle panel. There are $n$ hole-pockets in the center of the Brillouin zone (BZ) and at the $K$ ($K'$) points. SOC bands are shown in the bottom panel.
		\label{fig:bands}
	\end{figure*}
	
	\section{\label{sec:a2f}Mode-resolved spectral function $\alpha^2F(\omega)$}
	Different modes from different atomic species are contributing to the spectral function, $\alpha^2F(\omega)$. The mode-resolved $\alpha^2F(\omega)$ are illustrated in Fig. S2. As highlighted in the main text, it is evident that the primary contribution stems from NbSe$_2$ layers (Fig. S2(a,c,d,f-h)), while the contribution from the MoS$_2$ layer is minimal (Fig. S2(b,e,i)). This holds true across all three configurations. Notably, the outermost NbSe$_2$ layer stands out as the most significant contributor to  $\alpha^2F(\omega)$ among all layers.
	
	\begin{figure}[!htb]
		
		\begin{tabular}{l}
			\includegraphics[scale=0.7]{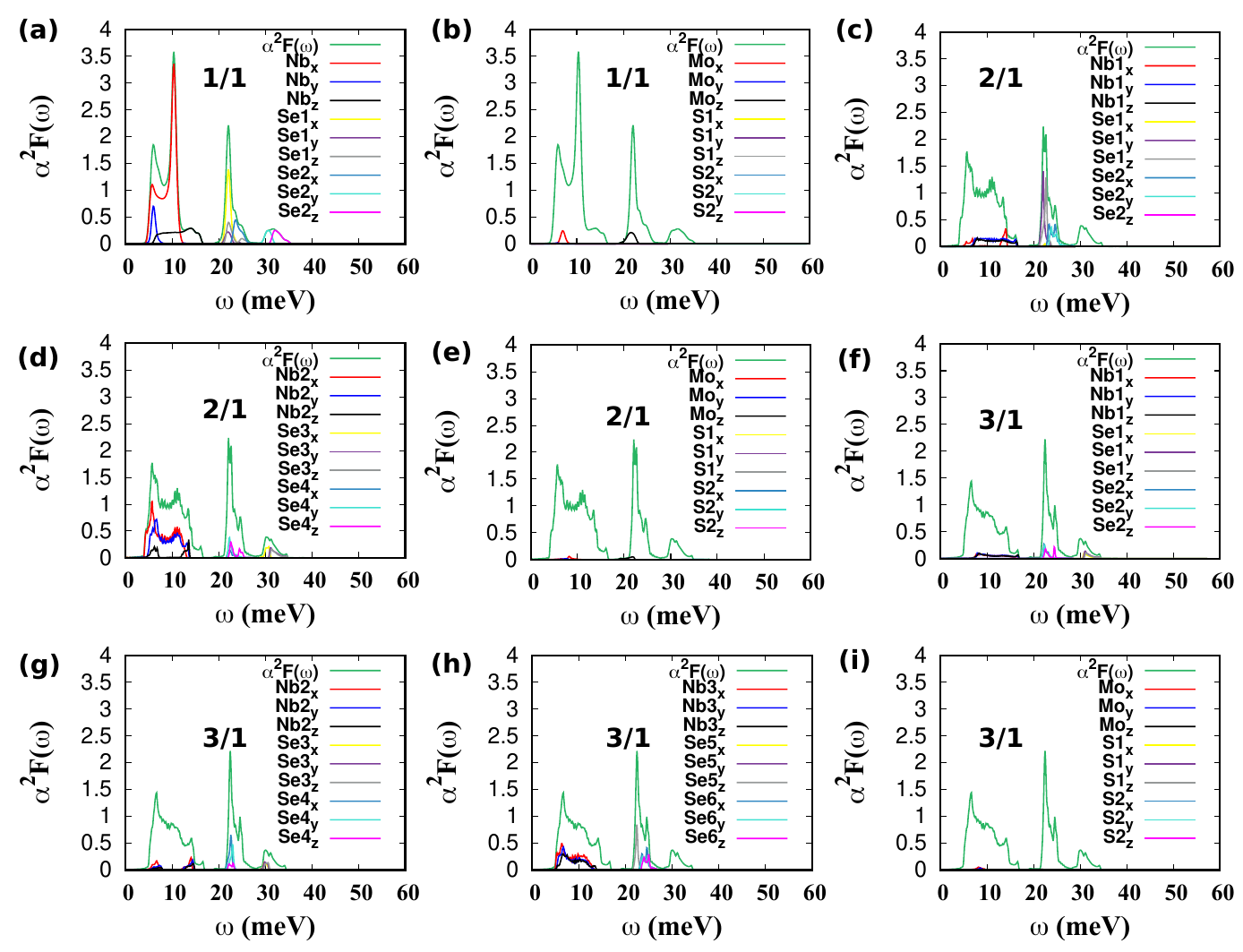}
		\end{tabular}
		\textbf{Fig. S2} Phonon mode-resolved spectral function at $\mu_{c}^* = 0.05$ for (a,b) 1/1, (c-e) 2/1 and (f-i) 3/1 configurations. Clearly, NbSe$_2$ layers are primarily responsible for the EPC.
		\label{fig:mode-a2f}
	\end{figure}
	
	\section{\label{sec:soc} Effect of spin-orbit coupling on $T_c$}
	Incorporating SOC diminishes the superconducting $T_c$, as seen in Fig. S3, where the temperature dependence of gap functions is shown. $T_c$ for 1/1 and 2/1 is 30K and 26.6K, respectively. With our computational resources we are limited up to 2/1 heterostructure for SOC calculations.
	
	\begin{figure}[!htb]
		
		\includegraphics[scale=0.75]{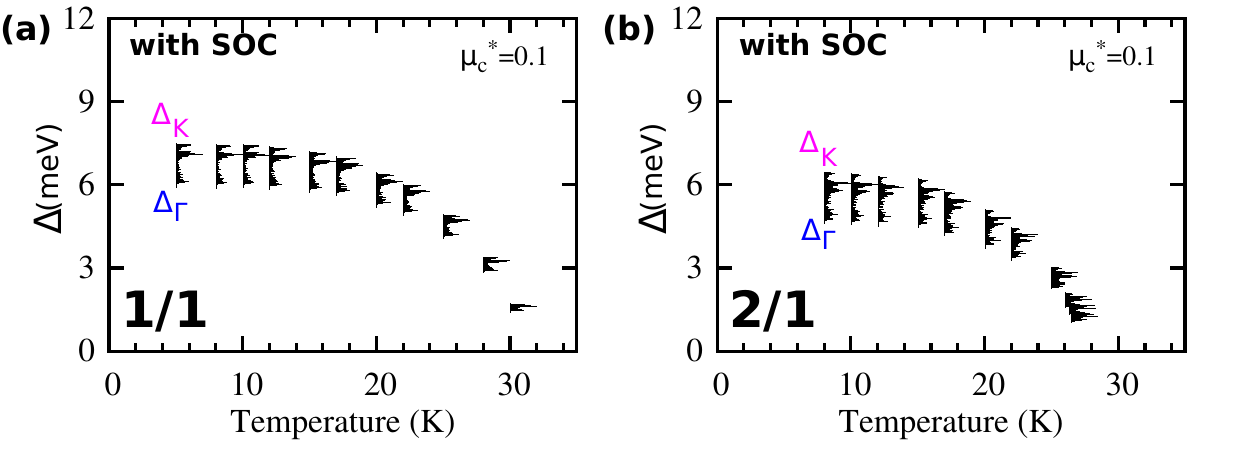}\\
		\textbf{Fig. S3} The energy distribution of superconducting gap ($\Delta$) as a function of temperature at $\mu_c^* = 0.1$ after including SOC for (a) 1/1 and (b) 2/1 heterostructures. A reduction in $T_c$ is clear in both the cases as compared to without SOC $T_c$.
		\label{fig:gap_soc}
	\end{figure}

\newpage

\textbf{AUTHOR DECLARATION}\\

\textbf{Code Availability}\\
Plotting scripts for the phonon-linewidth and Fermi surfaces are available from the corresponding author upon reasonable request.\\

	\bibliographystyle{unsrt}
	\bibliography{refs}